\documentclass[iop,numberedappendix]{emulateapj}
\usepackage{amsmath,graphicx}
\usepackage{bm}
\usepackage{times}
\newcommand{\Eq}[1]{Equation~(\ref{#1})}
\newcommand{\EQ}{\begin{equation}}
\newcommand{\EN}{\end{equation}}
\newcommand{\vv}{\mbox{\boldmath $v$} {}}
\newcommand{\uu}{\mbox{\boldmath $u$} {}}
\newcommand{\Sec}[1]{Section~\ref{#1}}

\newcommand{\St}{\text{St}}
\newcommand{\talp}{\alpha_5}
\newcommand{\teps}{\tilde{\epsilon}}
\newcommand{\tSt}{\text{\St}_2}
\newcommand{\halp}{\alpha_2}
\newcommand{\Stz}{\text{St}_0}
\newcommand{\rhoS}{\rho_{{\bullet}}}

\graphicspath{{.}{./Figures/}}

\date{\today,~ $ $Revision: 1.20 $ $}
\begin{document}

\title{Making Terrestrial Planets: High Temperatures, FU Orionis Outbursts, Earth, and Planetary System Architectures}
\shorttitle{Making Terrestrial Planets}

\author{Alexander Hubbard\altaffilmark{1}}
\altaffiltext{1}{Dept. of Astrophysics, American Museum of Natural History, New York, NY, USA}
\email{{\tt ahubbard@amnh.org}}

\begin{abstract}
Current protoplanetary dust coagulation theory does not predict dry silicate planetesimals,
in tension with the Earth. While remedies to this predicament have been proposed, they have generally
failed numerical studies, or are in tension with the Earth's (low, volatility dependent) volatile and moderately volatile elemental abundances.
Expanding on the work of \cite{2014ApJ...792L..27B}, we examine the implications of molten grain collisions and
find that they may provide a solution to the dry silicate planetesimal problem. Further, the source of the heating, be it
the hot inner disk or an FU Orionis scale accretion event, would dictate the location of the resulting planetesimals,
potentially controlling subsequent planetary system architectures. We hypothesize that systems which did undergo FU Orionis
scale accretion events host planetary systems similar to our own, while ones that did not instead host very close
in, tightly packed planets such as seen by Kepler.
\end{abstract}


\section{Introduction}

One of the more significant difficulties facing modern dust coagulation theory is that it seems to prohibit the straightforward formation
of dry (water poor) planetesimals or planets. That is, distinctly, in contradiction with clear observational
evidence both from within our Solar System and from extrasolar planetary systems \citep{marty2006water,2011ApJ...741...64K}.
The fundamental problem is that rocky material is not particularly sticky. As a consequence, the dust coagulation phase
for non-icy grains in protoplanetary disks is predicted to end long before the grains grow large enough to permit the formation of planetesimals.
While ice-rimmed grains are much stickier, potentially allowing for the formation of wet planetesimals \citep{2016SSRv..205...41B},
it is not clear how to make a dry planet from wet planetesimals.

Several potential ways around this difficulty have been proposed. We could be dramatically
underestimating the effective stickiness of silicates \citep{2015ApJ...812...67K,2016ApJ...832L..19A}. Similarly, dust grains could acquire sticky organic coatings
\citep{2000M&PSA..35Q..73H,2013EP&S...65.1159F},
increasing their stickiness above that of bare silicates.
Even though bouncing stalls growth, the Solar Nebula contained vast numbers of grains, some of which would
have been astoundingly lucky in their collisions \citep{2012A&A...544L..16W}.
Such lucky grains could potentially grow large enough to begin sweeping up small scale material, growing straight to planetesimal sizes.
Even if silicate growth is indeed prohibited, wet planetesimals formed outside the frost line could be scattered inwards, subsequently accreting local
dry dust. If that dry accretion proved sufficient to overwhelm their wet cores, those planetesimals could go on to
form dry planets.

In this paper we examine the possibility that high temperatures, which can dramatically increase the stickiness
of the dust, leads to dry planetesimal formation, as suggested by \cite{2014ApJ...792L..27B}.
As we will show, this route to dry planetesimal formation has interesting consequences for
planetary system architectures, 
and offers an ``in situ'' model for Mars' small size wherein planetesimal formation naturally ceases to operate efficiently
significantly outside of Earth's orbit \citep{2014Sci...344..479C}.
Taking this line further, it also suggests that there might
be two dominant initial planetary system architectures, one of which looks like our own, and the other more like systems with
tightly spaced inner planets \citep[or STIPs,][]{0067-0049-197-1-8,2014ApJ...792L..27B}.

\section{Statement of the difficulty}

The challenge bouncing poses to the formation of rocky planetesimals has been known since its introduction to dust coagulation theory \citep{2010A&A...513A..57Z}.
In a recent epic endeavor \cite{2016ApJ...818..200E} verified that the problem persists even taking into account lucky grains
\citep{2012A&A...544L..16W,2013ApJ...764..146G}.
In this section we quickly sketch how large modifications to our understanding would be required
to allow dry silicates to form planetesimals.

\subsection{Stokes numbers}

The dynamics of dust grains in a protoplanetary disk are controlled by their aerodynamical drag:
\EQ
\partial_t \vv = -\frac{\vv -\uu}{\tau}, \label{dvdt}
\EN
where $\vv$ is the dust grain's velocity, $\uu$ the gas velocity at the dust grain's location, and $\tau$ the drag based stopping time.
It is conventional to non-dimensionalize $\tau$ through the local Keplerian orbital frequency $\Omega_K$ to define a Stokes number:
\EQ
\St \equiv \tau \Omega_K. \label{St0}
\EN
In regions associated with the formation of the Earth, the dust grains we will consider are small enough to be in the Epstein drag regime, with 
midplane Stokes numbers
\EQ
\Stz = \frac{\pi a \rhoS}{2\Sigma_g}, \label{Stz}
\EN
where $a$ and $\rhoS$ are the dust grain radius and solid density, and $\Sigma_g$ the disk's gas surface density.

For the inner disk we will also consider solids larger than the gas mean-free-path, in the Stokes drag regime, 
which depends on the gas mean molecular mass rather than the gas density.
Such
grains have Stokes numbers
\EQ
\Stz = \frac 29 \frac{\rhoS a^2}{m_g} \frac{\sigma_g}{H}, \label{StS}
\EN
where $m_g \simeq 2.3$\,amu and $\sigma_g \simeq 2 \times 10^{-15}$ \citep{1970mtnu.book.....C}
 are the gas mean molecular mass and molecular collisional cross-section, respectively.
Further, 
\EQ
v_{th} = \langle v \rangle = \sqrt{\frac{8 k_B T}{\pi m_g}} \label{vthermal}
\EN
 is the local gas thermal speed and 
 \EQ
 H =\sqrt{\frac{\pi}{8}} \frac{v_{th}}{\Omega_K} \label{Heq}
 \EN
 defines the local vertically isothermal pressure scale height.

\subsection{Radial drift}
Due to their radial pressure gradient, disks orbit slower than Keplerian, with a velocity deficit $\delta v$. If the pressure is a power-law in radius
with power $s$, 
$\delta v$ is given by:
\EQ
\delta v = \frac{\pi }{16} s \frac{v_{th}^2}{R} \Omega_K^{-1} \simeq 50\text{ m s}^{-1}. \label{dv}
\EN
where $s \simeq 13/4$ for a Hayashi MMSN \citep{1981PThPS..70...35H}.
Note that  $\delta v$ is independent of $R$ in a disk with $T \propto R^{-1/2}$. A dust grain with $\St \ll 1$ drifts inwards at a speed
\EQ
v_r \simeq 2\, \St \, \delta v \simeq \St \times 10^4 \text{ cm s}^{-1}, \label{vr}
\EN
allowing us to define a drift time scale
\EQ
\tau_R = \frac{R}{v_r} \simeq \frac{R}{2 St \delta v} = \frac {1}{2 \pi s \, \St} \frac{R^2 }{H^2} \text{Orb}, \label{tR}
\EN
where $\text{Orb}$ is the local orbital period.
Local structures such as pressure bumps alter $\delta v$, potentially reversing its sign, allowing the trapping of particles \citep{2013ApJ...763..117D}.

\subsection{Streaming Instability}

The Streaming Instability (or SI) is a mechanism that produces planetesimals from the minimal constituent size, significant because
dust coagulation theory does not predict large dust grains. 
The trigger conditions
for the SI are nonetheless relatively large dust grains of $\tSt \equiv \St/0.01 \simeq 1$ combined with elevated local dust-to-gas mass ratios of
$\teps \equiv \epsilon/3 \times 10^{-2} \simeq 1$ \citep{2015A&A...579A..43C}. While that value is above the expected overall value of $\epsilon=5 \times 10^{-3}$
 for the Solar Nebula inside of the frost line \citep{2003ApJ...591.1220L}, there are several ways to effectively enhance $\epsilon$, including strong settling, or radial pressure 
 or temperature traps \citep{2013ApJ...763..117D,2016MNRAS.456.3079H}. 
 
\subsection{Bouncing Barrier}

Bouncing sets the upper size limit for locally coagulating reasonably compact (non-fractal) dust grains.
We assume that the critical
bouncing velocity $v_b$ is significantly below the fragmentation velocity, allowing us to neglect the latter.
Fragmentation velocities only modestly
above the bouncing velocity would further reduce the upper size limit.
The bouncing barrier and its consequences has been explored numerically
\citep[e.g.~][]{2010A&A...513A..57Z,2012A&A...544L..16W,2013ApJ...764..146G,2016ApJ...818..200E}, and found to prohibit
grain growth to SI triggering dust sizes.

From \cite{2010A&A...513A..56G}, the critical bouncing velocity
for porous dry silicates is
\begin{align}
v_b &\simeq f \times10^{-1} \left(\frac{m}{10^{-4}\text{ g}}\right)^{-5/18} \text{ cm s}^{-1}, \nonumber \\
&\simeq f \times 3 \times 10^{-4} \tilde{R}^{5/4} \tSt^{-5/6}  \text{ cm s}^{-1}, \label{vb}
\end{align}
where $m$ is the dust grain mass, $\tilde{R}$ the orbital position in a.u.,
and $f$ an arbitrary scaling factor which parameterizes the dust's surface parameter uncertainties.
While strict sticking/bouncing velocity cut-offs such as \Eq{vb} are simplifications, we can estimate the rate of sticking events
as the rate of collisions at velocities $v<v_b$.
In \Eq{vb}, following \cite{2010A&A...513A..56G}, the grains have volume filling factors of $\phi=0.12$ and monomer densities
of $\rho=2$\,g\,cm$^{-3}$, slightly at odds with the canonical molten solid density we will adopt later of $\rhoS = 3$\,g\,cm$^{-3}$ 
\citep{Friedrich2014}.

Assuming turbulent stirring, we can estimate the collision velocity scale as \citep{1980A&A....85..316V}
\EQ
v_0 \simeq \sqrt{\alpha \St}\, c_s = 38 \sqrt{\talp \tSt}\, \tilde{R}^{-1/4}\,  \text{ cm s}^{-1}. \label{v0}
\EN
In this section, to give dust coagulation the best chance, we assume that the disk is locally barely turbulent, approximating an
 $\alpha$ disk with $\talp \equiv \alpha/10^{-5} \simeq 1$, appropriate for quiescent dead zone midplanes \citep{2009ApJ...704.1239O}.
In the limit of $v_b \ll v_0$ we can estimate the rate at which a given grain collides with like grains \text{at velocities $v<v_b$} to be \citep{2016ApJ...826..152H}
\EQ
S \simeq \sqrt{2 \pi} n a^2 v_0 \left(\frac{v_b}{v_0}\right)^4, \label{S0}
\EN
where the number density of the grains is
\EQ
n = \frac{\epsilon \rho_g}{m}.
\EN
Equating the drift rate $v_r/R$ from \Eq{tR} with the estimated sticking rate $S$ from \Eq{S0} gives an extremely optimistic estimate for the maximum size dust grains can
reach before radially drifting out of our zone of interest. 
That becomes
\EQ
\sqrt{2 \pi} n a^2 v_0 \left(\frac{v_b}{v_0}\right)^4 = \frac{v_r}{R},
\EN
which reduces to
\EQ
f \simeq 6.6 \times 10^4 \times  \tSt^{41/24} \talp^{3/8} \teps^{-1/4} \tilde{R}^{-11/8}. \label{f0}
\EN
\Eq{f0} implies that our estimate of the critical bouncing velocity (Equation~\ref{vb}) would need to be low
by $4$ to $5$ orders of magnitude to allow the SI (i.e.~$\tSt \simeq 1$). Thus, dust coagulation theory predicts
that dry planetesimals will not form.

\subsection{Possible remedies}

That prediction is at odds with evidence from the Solar System. One obvious remedy
would be poor estimates of the sticking parameters of dry silicates \citep{2015ApJ...812...67K}, 
or of the appropriate monomer size \citep{2016ApJ...832L..19A}, but \Eq{f0} makes it clear however how extreme those mis-estimates
would need to be.
Magnetic interactions could also increase the sticking rates of dust grains, but \cite{2016ApJ...826..152H} showed that the critical
magnetic velocity is expected to be well below the collisional velocity $v_0$ from \Eq{v0} for $\St \gtrsim 10^{-2}$, so magnetic interactions
should die off well before the SI can be triggered.

Fischer-Tropsch-like processes could have lead to the synthesis of complex organic molecules
within the Solar Nebula \citep{2000M&PSA..35Q..73H,2013EP&S...65.1159F}, depositing
layers of sticky organic solids on the dust \citep{Kuga09062015}.
Those layers would have dramatically increased the dust's stickiness. However, the Earth is not merely
dry, but also strongly depleted in carbon. While the degree of that depletion is uncertain, it seems likely that adding more than just a few
mass percent of chondritic material would oversupply the Earth's entire carbon budget \citep{marty2013primordial},
to say nothing of even more strongly carbon rich organic gunk. Thus, any organic gunk model would need
to demonstrate that it can operate without oversuppling carbon. Indeed,
if recent laboratory work on catalyzing Fischer-Tropsch like processes is confirmed \citep{2016M&PS...51.1310N}, models for the formation of the Earth will
likely need to explain the destruction of such layers!

Lucky grains, which undergo a series of low probability low velocity sticking collisions,
can grow significantly larger than their more abundant staid counterparts, but numerical studies have
found that lucky grains are too rare to trigger planetesimal formation \citep{2016ApJ...818..200E}.
That also rules out importing large dust grains from outside
the frost line. Importing sufficient numbers of fully formed planetesimals from outside
the frost line would allow planet formation to proceed, but those planetesimals would need to accrete sufficient local
dry material to drop their bulk volatile abundances.

Thanks to the radial pressure gradient, planetesimals move with a speed of about $\delta v= 50$\,m\,s$^{-1}$ with respect to the local gas
and small scale dust (Equation~\ref{dv}). Assuming perfect, purely geometric sweep-up of ambient dust, those planetesimals would grow at a rate of
about
\EQ
\partial_t a = \frac{\epsilon \rho_g}{4 \rhoS}\delta v \simeq \frac 38 \tilde{R}^{-11/4} \text{ cm yr}^{-1}. \label{sweepup}
\EN
Thus, purely geometrically accreting planetesimals will only gain a few km of dry surface over a Myr, which seems unlikely to sufficiently overwhelm
their volatile rich core. Pebble accretion could speed the process as long as the available grains are comparable in size to chondrules \citep{2015SciA....115109J},
but pebble accretion's size sensitivity poses its own difficulties.

\section{High temperature dust collisions}

The purpose of this paper is to discuss how high temperatures could act as a mechanism for allowing dry dust to trigger the SI.
Dust grains collide and interact as liquids as long as the temperature of the solids is above about $1100$\,K \citep{2004M&PS...39..531C}.
In that regime, the grains
are much stickier and less prone to fragmentation as long as the grain sizes are small enough
and the temperature not too high. Such temperatures are associated with sufficient thermal ionization
of potassium to allow for full MRI activity \citep{1991ApJ...376..214B,1996ApJ...457..355G}; so we normalize through $\halp \equiv \alpha/0.01$ for high temperatures.
\cite{2014ApJ...792L..27B} suggested that the inner disk region hot enough to allow liquid grain collisions could have seen direct dust coagulation
to planetesimal sizes, producing STIPs-like systems. However, as we will show, molten collisions are not immune to bouncing and fragmentation.

\subsection{Liquid bouncing and splashing}
\label{bounce/splash}

Highly viscous molten grains collide as solids, rather than liquids.
As long as those effectively solid grains are not extremely sticky ($f \sim 6.6 \times 10^4$, Equation~\ref{f0}),
the bouncing barrier remains in place at high viscosity.
If the grains are instead barely viscous, and too large for surface tension
to contain the collisional energy, they will splash upon collision.
The viscosity condition for colliding as liquids is \citep{2004M&PS...39..531C,2015Icar..254...56H}:
\EQ
\eta \lesssim 6 \times 10^7 \left(\frac{a}{\text{cm}}\right) \left(\frac{v}{\text{cm s}^{-1}}\right)^{-1/5} \text{ P}, \label{etab}
\EN
where P stands for Poise, the cgs unit of dynamical viscosity.

While the condition for avoiding splashing is not certain, from \cite{2014ApJ...797...30J} we have:
\EQ
\eta \gtrsim 4 \times 10^{-12}  \left(\frac{a}{\text{cm}}\right)^{3} \left(\frac{v}{\text{cm s}^{-1}}\right)^{5} \text{ P}, \label{etas}
\EN
where we have assumed a surface tension of $400$\,dyn\,cm$^{-1}$ ($0.4$\,N\,m$^{-1}$)
for molten chondritic material \citep{2002ApJ...564L..57S}. Combining Equations~(\ref{etab}) and (\ref{etas}) we find the condition
\EQ
\left(\frac{a}{\text{cm}}\right) \left(\frac{v}{\text{cm s}^{-1}}\right)^{2.6} < 4 \times 10^{9} \label{visc_cond}
\EN
for there existing a viscosity for which liquid, non-splashing collisions are possible.

At temperatures near $T \simeq 10^3$\,K, the thermal speed is approximately $c_s \simeq 2 \times 10^5$\,cm\,s$^{-1}$ and
the turbulent collision speed is about (Eq~\ref{v0}):
\EQ
v \simeq 2 \sqrt{\St} \times 10^4 \text{ cm s}^{-1}, \label{vhot}
\EN
where we have assumed $\alpha \sim 10^{-2}$.  In a Hayashi MMSN \citep{1981PThPS..70...35H}, with $\Sigma_g = 1700\,\tilde{R}^{-1.5}$,
assuming $T \simeq 10^3$\,K,
we can combine Equations~(\ref{Stz}), (\ref{StS}), (\ref{visc_cond}), and (\ref{vhot}) to find the largest possible grain sizes:
\begin{align}
&a_E \lesssim 5.7\, \tilde{R}^{-0.85} \text{ cm},\label{aE}\\
&a_S \lesssim 6.3\, \tilde{R}^{0.54} \text{ cm}, \label{aS}
\end{align}
for the Epstein (dominant outwards of $0.93$\,au) and Stokes (dominant inwards of $0.93$\,au) drag regimes, respectively.

Equations~(\ref{aE}) and (\ref{aS}) imply largest possible Stokes numbers of
\begin{align}
&\St_E \lesssim 0.016\, \tilde{R}^{0.65}, \\
&\St_S \lesssim 0.015\, \tilde{R}^{-0.42},
\end{align}
respectively. Those sizes require viscosities of
\begin{align}
&\eta_E \simeq 7.7 \times 10^7 \tilde{R}^{-0.9} \text{ P}, \\
&\eta_S \simeq 8.5 \times 10^7 \tilde{R}^{0.6} \text{ P},
\end{align}
to reach. Thus, the liquid bouncing/fragmentation barriers
prevent growth to $\St \gg 10^{-2}$, with critical viscosities to reach $\St \simeq 10^{-2}$ of $\eta \sim 5-10 \times 10^7$\,P.
The precise viscosities of Solar Nebula solids are unknown, but those values would have required temperatures near $T=10^3$\,K \citep{2015Icar..254...56H}.

Being molten is not a panacea for the formation of terrestrial planets, and
the collision of heated solids does not allow direct coagulation to $\St \gg 10^{-2}$. However, it does seem
likely that temperatures in a relatively narrow range near $T \sim 10^3$\,K permit nearly perfect sticking up to $\St \sim 10^{-2}$,
conspicuously the critical value for triggering the SI.
Unfortunately, we lack detailed laboratory experiments measuring the effective viscosity of chondritic melts of centimeter grains
at the relevant temperatures,
with thermal histories, and we lack detailed zero-g, low pressure splashing experiments. That limits how thoroughly
molten grain growth can be modeled beyond the simple estimates above.

\subsection{Growth rates and comparison to radial drift}

Neglecting bouncing and fragmentation,
the growth rate of a given dust grain is the same as sweep-up up to the details of the velocity (Equation~\ref{sweepup}):
\EQ
\frac{\partial a}{\partial t} =  \frac{\epsilon \rho_g v}{4 \rhoS}. \label{dadt0}
\EN
From Equations~(\ref{Stz}) and (\ref{StS}) we can see that the Stokes number's quadratic dependence on $a$ means that 
the Stokes number increases more rapidly
in the Stokes drag regime, and we need only consider the growth time scales in the Epstein regime where 
\EQ
St  = \sqrt{\frac{\pi}{8}} \frac{a \rhoS \Omega_K}{\rho_g v_{th}}. \label{St}
\EN
gives the Stokes number at all altitudes.

Combining Equations~(\ref{dadt0}) and (\ref{St}) we find
\EQ
\frac{\partial \St}{\partial t} =\sqrt{\frac{\pi^3}{32}}  \epsilon \left(\frac{v}{v_{th}}\right) {\text{Orb}}^{-1} \simeq  \epsilon \left(\frac{v}{v_{th}}\right) {\text{Orb}}^{-1}.
\label{dStdt}
\EN
We can insert \Eq{v0} into \Eq{dStdt} to find
\EQ
2 \sqrt{\St(t)}  = \sqrt{\alpha} \epsilon \frac{t}{\text{Orb}} + 2 \sqrt{\St(0)}. \label{Stt0}
\EN
Neglecting $\St(0)$ we arrive at
\EQ
\St(t) \simeq \frac{\alpha \epsilon^2}{4} \left(\frac{t}{\text{Orb}}\right)^2 = 2.25 \times 10^{-6} \,\halp\, \teps^2 \left(\frac{t}{\text{Orb}}\right)^2, \label{Stt}
\EN
or alternatively, the time to reach a given $St$ is
\EQ
t(\St) \simeq \frac{2}{\epsilon} \sqrt{\frac{\St}{\alpha}} \text{Orb} \simeq 67 \frac{1}{\teps} \sqrt{\frac{\tSt}{\halp}} \text{Orb}, \label{tSt}
\EN
where we recall that $\alpha$ and $\St$ are both normalized to $10^{-2}$:
$\tSt=\St/10^{-2}$ and $\halp=\alpha/10^{-2}$.

Comparing Equations~(\ref{tR}) to (\ref{tSt}) we find that dust growth outpaces radial drift as long as
\EQ
 \St <  \left[\frac {\alpha \epsilon^2}{16 \pi^2  s^2} \frac{R^4 }{H^4}\right]^{1/3} \simeq 0.16\,\halp^{1/3} \teps^{2/3} \tilde{R}^{-1/3},
 \label{radial_drift_limit}
 \EN
Dust growth can outpace radial drift up to SI triggering dust grain sizes
for reasonable dust-to-gas mass ratios even for modest turbulence and relatively thick disks.

\section{Sources of heating and planetary system architecture}

\subsection{Inner disk}

\cite{2014ApJ...792L..27B} proposed the inner disk, with $T \sim 1500$\,K, as a region where partially molten grains
could coagulate well beyond the conventional bouncing and fragmentation limits, envisioning coagulation up to planetesimal sizes. As we showed in \Sec{bounce/splash}
though, new versions of the bouncing and fragmentation barriers show up
as dust grains grow beyond about $\St \simeq 0.01$. However, that approximate Stokes number of $\St=0.01$ is sufficiently large to allow
the SI \citep{2007Natur.448.1022J} to trigger, albeit in a narrow radial annulus where the temperature is modestly above $T=10^3$\,K and the 
corresponding viscosity
of the molted grains several times $10^7$\,P. That would naturally occur at around $R=0.1$\,au,
and the temperature range is particularly interesting as it is close to the temperature required for sufficient thermal
ionization to allow MRI activity \citep{1996ApJ...457..355G}. Thus, the region should occur near the inner edge of the dead zone, a location theorized to concentrate dust
particles \citep{2009A&A...497..869L}.

We have then the potential for rapid planetesimal formation through the SI if that concentrated dust
can be brought into regions of the correct temperature. That would occur either if
those temperatures are the normal background temperature at the edge of the dead zone,
or through secular evolution of the disk's radial profile.
Thus, the picture of \cite{2014ApJ...792L..27B}, with terrestrial planetesimals forming in the hot inner disk,
survives in a modified fashion. Significantly, it can potentially explain the recent observation of systems with
tightly spaced inner planets (or STIPs) representing a significant fraction (more than $10$\%) of stellar systems.

\subsection{FU Orionis type events}

However, that inner disk picture is a poor match for the formation of the Earth, and due to the large amount of migration that would be required,
Solar System like planetary architectures in general. The Earth shows a smooth volatility dependent
depletion pattern across a broad condensation temperature range of $T \sim 700-1400$\,K \citep{2000SSRv...92..237P,2003TrGeo...2..547M}, which is ill-fit
by planetesimal formation in a narrow temperature range. \cite{2014Icar..237...84H} put forth a model to explain the Earth's abundance pattern
through time-dependent heating and cooling from an FU Orionis type event in the early Solar System \citep{1996ARA&A..34..207H}. 

FU Orionis type events are dramatic, long lived accretion events, associated with luminosity increases of $4-6$\,magnitudes and
time scales of $50-100$\,yr \citep{1996ARA&A..34..207H,JOEL16}.
FU Orionis events (or FUors) can raise the temperature at Earth's orbital position to about $1350$\,K \citep{2014Icar..237...84H},
and drive the frost line out to about $40$\,au \citep{2016arXiv160703757C}. An FUor that strong
would raise the temperature above $1000$\,K out to Mars' orbit, while a weaker, $4$ magnitude FUor would still raise the temperature above $1000$\,K out
to about Venus' orbit.
A long lived, $100$\,yr life time FUor outburst can satisfy the time scale constraint from \Eq{tSt} out to a bit past Earth's orbit,
while a $50$\,yr FUor would still satisfy \Eq{tSt} out to Venus'.
While the frequency
of FU Orionis events is not yet certain, statistics are consistent with a large fraction of protoplanetary disks undergoing one or several
FU Orionis outbursts \citep{1996ARA&A..34..207H}.

We therefore propose that molten collisions during an FU Orionis event were the solution to the bouncing and fragmentation barriers
for dry silicates at the locations of the terrestrial planets in our Solar System. Thermal processing and molten
collisions could explain both the existence of dry planetesimals at Earth's position, and
the Earth's volatile abundance trend \citep{2014Icar..237...84H}. The scenario suggests an ``in situ'' explanation for Mars' small size \citep{2014Sci...344..479C}:
the temperature and growth time scales
required mean that the mechanism ceases to operate somewhere between Earth and Mars' orbital positions (see Equation~\ref{tSt}),
naturally reducing the number density of dry planetesimals beyond Earth's orbit. By cycling the temperature high
enough to burn off organic surface layers, the model would also help explain why the Earth is carbon-poor even
if carbon deposition is expected \citep{marty2013primordial,2016M&PS...51.1310N}.
In our picture, molten collisions and evaporation/recondensation mechanisms \citep{2017MNRAS.465.1910H}, 
rather than conventional dust coagulation, put the Solar System's architecture
in place very early in the Solar Nebula's existence.
In addition to terrestrial planets,
the duration and heating of an FUor leads to evaporation/recondensation cycles that might have driven icy planet formation in the outer disk \citep{2017MNRAS.465.1910H}.

The briefer (up to a few year duration) accretion events associated with Ex Lupi and Exors in general \citep{2007AJ....133.2679H}
are too brief for any given eruption to allow molten collisions to drive significant growth far beyond the inner edge of the dead zone, and are generally
weaker than FUors.
If Exors survive as episodic accreters long enough to accumulate decades of integrated outburst time, then they might drive intermittent molten dust growth to SI triggering sizes.
Note however that dust drifts radially during quiescent periods as well as during active ones, so
 the radial drift condition (Equations~\ref{radial_drift_limit}) becomes more stringent.

\section{Discussion and Conclusions}

Thermal processing of dust and molten grain collisions at $T \gtrsim 10^3$\,K provides dry silicate dust a way past the bouncing and fragmentation barriers 
\citep{2010A&A...513A..57Z} to the Streaming Instability (or SI), although it likely does not permit direct coagulation to Stokes numbers $\St \gg 10^{-2}$.
 Significantly, most other mechanisms that have been proposed
to bypass those barriers that have not been ruled out by numerical study \citep{2016ApJ...818..200E} are inconsistent
with the volatile abundances of the terrestrial planets in the Solar System \citep{2013EP&S...65.1159F,marty2013primordial}.
This thermal processing seems likely to take two forms depending on the host disk dynamics, and those forms would set the planetary system
architecture in an \emph{in situ} planet formation scenario.

The first form, suggested by \cite{2014ApJ...792L..27B}, appeals to the inner disk where temperatures are naturally elevated.
This region is also attractive due to being close to the inner edge of the dead zone, a location prone
to concentrating solid material \citep{2009A&A...497..869L}. This form is inconsistent with the composition
of the Earth \citep{2000SSRv...92..237P,2003TrGeo...2..547M}, but is attractive for explaining
the systems with
tightly spaced inner planets (or STIPs) seen by Kepler in a large fraction of stellar systems \citep{0067-0049-197-1-8}.

The second form appeals to the century scale \citep{1996ARA&A..34..207H,JOEL16} heating from an FU Orionis
scale accretion event, which would provide the required heating  for the required time out to an orbital location between Venus' orbit and Mars'.
The heating time scale and resulting radial planetesimal distribution would be consistent with Earth's composition \citep{2014Icar..237...84H} and with Mars' small size \citep{2014Sci...344..479C};
and appeals to a process that we have proposed to promote the formation of Jupiter and Saturn-like gas giants \citep{2017MNRAS.465.1910H}.
This suggests that stellar systems which hosted FU Orionis outbursts would tend to host planetary systems like our own, with
full-sized dry terrestrial planets in an orbital band extending from the inner edge of the dead zone into the habitable zone.
FUors naturally lead to planetesimal formation through the inner habitable zone (Venus to Earth). Mar's size suggests that
modestly sized planets can still form too far out for immediate planetesimal formation via molten grain collisions, fully populating the habitable zone
with planets.
Thus, we next predict a planet gap extending from the outskirts of the habitable zone to the frost line.
Outside the frost line, the accretion events would have trigged gas giant formation.

The trigger mechanism(s) for FU Orionis scale accretion events are not yet known, and given their long
duty cycles, those mechanism(s) will remain uncertain for the foreseeable future. However,
they appear to be order-unity events in protostellar evolution \citep{1996ARA&A..34..207H}, while STIPs are order-unity outcomes of planet formation.
If STIPs and FU Orionis hosting systems are mutually exclusive, as we have suggested, then FU Orionis mechanisms would have to divide protostellar systems into
two roughly even categories. This is not unreasonable: as an example, Hall MRI \citep{2001ApJ...552..235B} depends
on whether the background magnetic field is aligned with or anti-aligned with the protoplanetary disk's rotation.
It is unclear which orientation would be more likely to lead to massive accretion events, but numerical simulations suggest that the difference between
aligned and anti-aligned systems could
easily determine the fate of the disk \citep{2014ApJ...791..137B}. The idea that planetary system architectures may ultimately depend
on something as innocuous as the chance relative orientation of the ambient magnetic field and their protoplanetary disk's rotation is intriguing. Future observations
of the relative rates of planetary system architectures will provide us with a new window on protoplanetary disk dynamics, including
constraints on the trigger mechanism for massive accretion events.

\section*{Acknowledgements}
The research leading to these results was funded by NASA OSS grant NNX14AJ56G. Mordecai-Mark Mac Low, Denton S. Ebel, and the anonymous referee
provided advice, assistance, and valuable suggestions for improvements to the manuscript.

\bibliography{Terrestrial}

\end{document}